\UseRawInputEncoding
\documentclass[superscriptaddress, notitlepage, reprint]{revtex4-1}
\usepackage[english]{babel}
\usepackage{amsmath,amsthm}
\usepackage{amsfonts}
\usepackage[pdfborder={0 0 0}, colorlinks=true, urlcolor=blue, linkcolor=blue, citecolor=blue]{hyperref}
\usepackage{color}
\usepackage{graphicx}
\usepackage{float}
\usepackage{subfigure}
\usepackage{lipsum}
\usepackage{epstopdf}
\usepackage{dcolumn}
\begin{document}

\title{Quantum phases transition revealed by the exceptional point in Hopfield-Bogoliubov matrix}
\author{Dong  Xie}\email{xiedong@mail.ustc.edu.cn}
\affiliation{College of Science, Guilin University of Aerospace Technology, Guilin, Guangxi 541004, People's Republic of China}
\affiliation{State Key Laboratory for Mesoscopic Physics, School of Physics, Frontiers Science Center for Nano-Optoelectronics, and
Collaborative Innovation Center of Quantum Matter, Peking University, Beijing 100871, People's Republic of China}
\author{Chunling Xu}
\affiliation{College of Science, Guilin University of Aerospace Technology, Guilin, Guangxi 541004, People's Republic of China}
\author{An Min Wang}
\affiliation{Department of Modern Physics, University of Science and Technology of China, Hefei, Anhui 230026, People's Republic of China}

\begin{abstract}
We use the exceptional point in Hopfield-Bogoliubov matrix to find the phase transition points in the bosonic system. In many previous jobs, the excitation energy vanished at the critical point. It can be stated equivalently that quantum critical point is obtained when the determinant of Hopfield-Bogoliubov matrix vanishes. We analytically obtain the Hopfield-Bogoliubov matrix corresponding to the general quadratic Hamiltonian. For single-mode system the appearance of the exceptional point in Hopfield-Bogoliubov matrix is equivalent to the disappearance of the determinant of Hopfield-Bogoliubov matrix. However, in multi-mode bosonic system, they are not equivalent except in some special cases. For example, in the case of perfect symmetry, that is, swapping any two subsystems and keeping the total Hamiltonian invariable, the exceptional point and the degenerate point coincide all the time when the phase transition occurs. When the exceptional point and the degenerate point do not coincide, we find a significant result. With the increase of two-photon driving intensity, the normal phase changes to the superradiant phase, then the superradiant phase changes to the normal phase, and finally the normal phase changes to the superradiant phase.
\end{abstract}
\maketitle

\section{Introduction}
Quantum phase transitions are playing an increasingly important role in many fields, such as, quantum metrology\cite{lab1,lab2,lab3,lab4,lab5,lab6,lab7,lab8,lab9,lab10,lab11,lab12}, which involves using quantum resources to improve measurement precision. The superradiant phase (SP) transition is one of the most important quantum phase transition, which was proposed in the Dicke model for the first time in 1970's\cite{lab1a}. In such model, a state is recognized as the normal phase (NP) when the cavity field is in the vacuum and the atoms are in their ground states; a state is recognized as the SP when the cavity field is intensely populated with two degenerate ground states and the atoms are excited simultaneously. The NP (SP) can be revealed by the order parameter of $\langle a\rangle=0$\cite{lab2a} ($\langle a\rangle\neq0$), where $a$ denotes the annihilation operator of the cavity mode.

In Ref.\cite{lab3a,lab15,lab16,lab17}, the critical point is revealed by the degenerate ground state. However, whether the degenerate ground state can be the general criteria for the occurrence of phase transition is a question worth exploring carefully.

In non-Hermitian system,  purely real eigenvalues of non-Hermitian Hamiltonian are obtained in the case of  parity-time (PT) unbroken phase; the eigenvalues become imaginary in the case of PT-broken phase\cite{lab17a}. The exceptional points (EPs) separate the PT-unbroken phase and the PT-broken phase. Both the eigenvalues and the eigenvectors coalesce at the EPs.

In the process of Hopfield-Bogoliubov transformation, there is a Hopfield-Bogoliubov (HB) matrix, which is non-Hermitian matrix generally. Like the PT symmetrical non-Hermitian Hamiltonian, the eigenvalues of HB matrix can be real and imaginary. We define EPs as the separating points between real and imaginary eigenvalues.
When one of the eigenvalues of HB matrix is 0, the ground state of system becomes degenerate. We define it as a degenerate point (DP).

In this article, we show that the EPs in HB matrix can reveal the phase transition in the bosonic system. The DP and the EP coalesce in the case of a perfectly symmetric system or a single-mode linear bosonic system. For multi-mode bosonic system, the DP cannot be the SP transition point in many cases.
Especially, in the two-mode bosonic system without counterrotating-wave interaction, the significant unconventional results are found that with the increase of two-photon driving intensity, the NP changes to the SP, then the SP changes to the NP, and finally the NP changes to the SP.

This article is organized as follows. In Section II, we obtain the general HB matrix for the multi-mode linear bosonic system and the corresponding EP. In Section III, we show that the EP and the DP coincide all the time for the single-mode linear bosonic system. In Section IV, the EP and the DP do not always coincide for the two-mode linear bosonic system due to the two-photon driving and counterrotating-wave  interaction. In Section V, the EP in three-mode quantum Rabi system is used to improve the estimation precision. We make a brief conclusion and outlook in Section VI.

\section{Hopfield-Bogoliubov matrix}
For a general multi-mode linear bosonic system composed of $N$ subsystems, the total Hamiltonian is quadratic, which can be described as
\begin{align}
H=\sum_{n=1}^NH_n+\sum_{i=1,i<j}^NH_{ij},
\label{eq:1}
\end{align}
in which,

\begin{align}
H_n=\omega_na^\dagger_n a_n+(\chi_na_n^2+h.c.),\\
H_{ij}=g_{ij}a_ia_j+\lambda_{ij}a_ia_j^\dagger+h.c.,
\label{eq:2}
\end{align}
where $H_{n}$ denotes the Hamiltonian for the $n$th subsystem  with $n=\{1,...,N\}$, $\omega_n$ is the resonance frequency of the bosonic
subsystem with the annihilation operator $a$ and the creation operator $a^\dagger$, $|\chi_n|$ denotes the strength of two-photon driving, and $\lambda_{ij}$ ($g_{ij}$) denotes the coupling strength of the rotating (counterrotating)-wave  interaction between the two subsystems.

By using an HB transformation\cite{lab13,lab14}, for the NP, the total Hamiltonian can be rewritten as a diagonal form
\begin{align}
H=\sum_{n=1}^N\Omega_n A_n^\dagger A_n+E_g,
\label{eq:3}
\end{align}
where the collective bosonic mode operators $A_n=\sum_{i=1}^N(\mu_{ni}a_{i}+\nu_{ni}a_{i}^\dagger)/\sqrt{\sum_{i=1}^N(|\mu_{ni}|^2-|\nu_{ni}|^2)}$ and $E_g$
represents the ground state energy. $A_n$ satisfies the commutation relation: $[A_i,A_j^\dagger]=\delta_{ij}$.
The coefficient vectors $(\mu_{n1},...,\mu_{nN},\nu_{n1},...,\nu_{nN})^T$ are eigenvectors of the HB matrix \textbf{M} which is derived by the commutation relation $[A_n,H]=\Omega_nA_n$
 \[
 \mathbf{M}= \left(
\begin{array}{ll}
\ \ \mathbf{A}\ \ \ \ \ \ \ \mathbf{B}\\
-\mathbf{B}^* \ \ -\mathbf{A}^*\\
  \end{array}
\right )  ,\tag{5}\label{eq:5}\]
with submatrix
 \[
 \mathbf{A}= \left(
\begin{array}{ll}
\omega_1\ \lambda_{12}\ \ \ldots\ \lambda_{1N}\\
\lambda_{12}^*\ \omega_2\ \ldots\  \lambda_{2N}\\
\ \vdots\ \ \ \ \ \vdots\ \ \ \ddots\ \ \ \vdots\\
\lambda_{1N}^*\ \lambda_{2N}^*\ \ldots\ \omega_N
 \tag{6}\label{eq:6} \end{array}
\right )  ,\]
 \[
 \mathbf{B}= \left(
\begin{array}{ll}
-2\chi_1\ -g_{12}\ \ \ldots\ -g_{1N}\\
-g_{12}\ \ -2\chi_2\ \ldots\  -g_{2N}\\
\ \ \ \vdots\ \ \ \ \ \ \ \vdots\ \ \ \ \ \  \ddots\ \ \ \vdots\\
 -g_{1N}\ -g_{2N}\ \ldots\ -2\chi_N
  \end{array}
\right )  .\tag{7}\label{eq:7}\]

The spectral values of the HB matrix $\textbf{M}$ have positive and negative symmetries due to the symmetry of the HB matrix: $\mathcal{C}\mathbf{M}\mathcal{C}^{-1}=-\mathbf{M}$, where $\mathcal{C}$ is described by
 \[
\mathcal{C}= \left(
\begin{array}{ll}
\ \ \textbf{0}\ \ \ \ \ \ \ \mathbf{I}\\
-\mathbf{I} \ \ \ -\mathbf{0}\\
  \end{array}
\right )  .\tag{8}\label{eq:8}\]
Even if all of the frequency values $\omega_n$ are greater than zero, we need to emphasize that $\Omega_n$ is not necessarily a positive eigenvalue of the HB matrix $\textbf{M}$. It can be obtained rigorously by mapping relationships \begin{align}
\Omega_n|_{\chi_n\rightarrow0,g_{ij}\rightarrow0,\lambda_{ij}\rightarrow0}\longrightarrow\omega_n.
\tag{9}\label{eq:9}
\end{align}

When the determinant of the HB matrix is equal to 0 ($Det(\textbf{M})=0$), the ground state will become degenerate. In Ref.\cite{lab15,lab16,lab17}, DPs are treated as the quantum critical point due to the excited energy becoming 0. We will show that the DP is not a critical point in many cases.

There are a lot of works on PT symmetric non-Hermitian Hamiltonian\cite{lab18,lab19,lab20,lab21,lab22,lab22a}, which  exists the EPs separating real and imaginary eigenenergy values.  At the EPs, eigenvalues and their corresponding eigenvectors coalesce. Similarly, due to that a general HB matrix $\textbf{M}$ is non-Hermitian and symmetrical,  there are also EPs.

For a closed system, the values of $\omega_n$ are real, $\mathbf{A}$ is Hermitian: $\mathbf{A}=\mathbf{A}^\dagger$. For special conditions, there are no counterrotating-wave interaction and two-photon driving:  $\mathbf{B}=0$. In this case, the HB matrix $\mathbf{M}$ is Hermitian. If at least one of the eigenvalues is equal to zero, then DP exists. For a Hermitian system, there is no imaginary eigenvalue, meaning that there is no EP. In this case, the absence of  phase transition indicates that the DP is not the same as the critical point.

Quantum phases transition can be revealed by the EPs. When the eigenvalues of the HB matrix is real, the system is in the NP; when the eigenvalues of the HB matrix is imaginary, the system is in the SP; the EPs denotes the transition point between the normal phase and the superradiance phase.

\section{single mode bosonic system and quantum Rabi system }
In this section, we will show that the EP and the DP coincide all the time for
the single-mode linear bosonic system. And make a corresponding comparison with the single-mode Rabi system.

For a single-mode linear bosonic system, the general Hamiltonian can be described as
\begin{align}
H_1=\omega a^\dagger a+(\chi a^2+h.c.)/2,\tag{10}
\label{eq:10}
\end{align}
The corresponding HB matrix is obtained from Eq.~(\ref{eq:5})
\[
 \mathbf{M_1}= \left(
\begin{array}{ll}
\omega\ \ \ -\chi\\
\chi^* \ \ -\omega\\
  \end{array}
\right )  .\tag{11}\label{eq:11}\]
The eigenvalues of $\mathbf{M_1}$ are $\pm\sqrt{\omega^2-|\chi|^2}$. As the way of above section,
$\Omega_1=\sqrt{\omega^2-|\chi|^2}$. Obviously, both the EP and the DP occur when $|\omega|=|\chi|$. Namely, in single-mode bosonic system, the DP can reveal the quantum phase transition due to that the EP and the DP appear at the same time.
The collective bosonic mode operator can be obtained $$A_1=\frac{\chi a+(\omega-\Omega_1)a^\dagger}{\sqrt{|\chi|^2-(\omega-\Omega_1)^2}}=\exp(i\theta)U(\xi)aU^\dagger(\xi),$$ where the phase is given by $\exp(i\theta)=\frac{\chi}{|\chi|}$ and the unitary squeezing operator is  $U(\xi)=\exp(\frac{\xi }{2}a^2-\frac{\xi^*}{2}{a^{\dagger2}})$, with the squeezing parameter defined as $\xi=\exp(-i\theta)\ln\frac{|\chi|+\omega-\Omega_1}{\sqrt{|\chi|^2-(\omega-\Omega_1)^2}}$.
The squeezing parameter $\xi$ diverges at the EP.

 The ground state is given by $|\psi_1\rangle=\exp(-\frac{\xi }{2}a^2+\frac{\xi^*}{2}{a^{\dagger2}})|0\rangle$. Performing measurements on the ground state of the system $|\psi_1\rangle$, we can obtain the values of $\{\theta, \omega, |\chi|\}$. The optimal estimation precision is given by the quantum Cram\'{e}r-Rao (CR) bound\cite{lab23,lab24,lab25}: $\delta^2\varphi\geq(\nu\mathcal{F_\varphi})^{-1}$, where $\nu$ denotes the total number of experiments and $\mathcal{F}_\varphi$ is the quantum Fisher information about the parameter $\varphi$ ($\varphi=\{\theta,\omega,|\chi|\}$). For the pure ground state, the quantum Fisher information $\mathcal{F_\varphi}$ can be achieved by $\mathcal{F_\varphi}=4[\langle\partial_\varphi\psi|\partial_\varphi\psi\rangle-|\langle\partial_\varphi\psi|\psi\rangle|^2]$.
When close to the EP, the dominant terms of the quantum Fisher information are
\begin{align}
\mathcal{F_\theta}\sim(\ln\Omega_1)^2/4,\tag{12}\label{eq:12}\\
\mathcal{F_\chi}\sim\frac{|\chi|^2}{4\Omega_1^4},\tag{13}\label{eq:13}\\
\mathcal{F_\omega}\sim\frac{|\omega|^2}{4\Omega_1^4}\tag{14}\label{eq:14}.
\end{align}
From above equations, we can see that according the quantum CR bound, the estimation  uncertainty of parameters $\theta,\omega,|\chi|$ is close to 0 ($\Omega_1\rightarrow0$) at the EP or the DP. It means that divergent quantum Fisher information can reveal the emergence of critical points (EP or DP). In addition, $\theta$ is independent of the phase transition point ($\Omega_1=\sqrt{\omega^2-|\chi|^2}=0$). As a result, the  scale of $\mathcal{F_\theta}$ is smaller than the scale of $\mathcal{F_\chi}$ and $\mathcal{F_\omega}$ near the critical point ($-\ln\Omega_1\ll\frac{1}{\Omega_1^2}$).

In the quantum Rabi system, the Hamiltonian can be described as
$H_R=\omega_0 a^\dagger a+\Delta/2\sigma_z+\eta(a^\dagger+a)\sigma_x$, where the pauli operator $\sigma_z=|e\rangle\langle e|-|g\rangle\langle g|$. In the limit $\omega/\Delta\rightarrow0$, using a Schrieffer-Wolff transformation\cite{lab26} and projecting onto the ground state $|g\rangle$, we can obtain a single mode bosonic system
\begin{align}
H_R'=(\omega_0-\frac{\eta^2}{2\Delta})a^\dagger a-\frac{\eta^2}{4\Delta}(a^2+{a^{\dagger2}})\tag{15}\label{eq:15}.
\end{align}
Let's redefine $\omega=\omega_0-\frac{\eta^2}{2\Delta}$ and $\chi=-\frac{\eta^2}{2\Delta}$, Eq.~(\ref{eq:15}) becomes Eq.~(\ref{eq:10}).  Ref.~\cite{lab26} showed that the normal phase and the superradiant phase are separated by the critical point $\eta/\sqrt{\omega_0\Delta}=1$. The critical point $\eta/\sqrt{\omega_0\Delta}=1$ also is the EP or the DP $|\omega|=|\chi|$.
 When all eigenvalues of the HB matrix are real ($\eta/\sqrt{\omega_0\Delta}<1$), the system is in the NP; when at least one of the eigenvalues is imaginary ($\eta/\sqrt{\omega_0\Delta}>1$), the system is in the SP.

\section{two-mode bosonic system and quantum Rabi system }
 In this section, we will demonstrate that the EP and the DP do not always coincide. Unusual phase transitions will be revealed.
\begin{figure}
\includegraphics[scale=0.56]{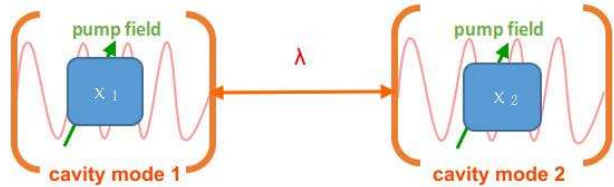}
\caption{\label{fig.1}The setup of two coupled cavity systems. The two cavity modes (red) couples with each other by the rotating-wave interaction: $\lambda a_1a_2^\dagger+h.c$, where $\lambda$ denotes the strength of resonant coupling. The cavity mode 1 (2) with angular frequency $\omega_{c1}$ ($\omega_{c2}$) is generated by a strong pump field (green) with angular frequency $2\omega_{d1}$ ($2\omega_{d2}$) via the second-order nonlinearity. $\chi_1$ and $\chi_2$ represent the strength of two-photon driving.}
\end{figure}

For two-mode linear bosonic system, the general Hamiltonian can be given by
\begin{align}
H_2&=\omega_1 a^\dagger_1 a_1+\omega_2 a^\dagger_2 a_2\nonumber\\
&+(\chi_1 a^2_1+\chi_2 a^2_2+ga_1a_2+\lambda a_1a_2^\dagger+h.c.),\tag{16}\label{eq:16}
\end{align}
The corresponding HB matrix is obtained
\[
 \mathbf{M_2}= \left(
\begin{array}{ll}
\omega_1\ \ \ \lambda\ \ \ -2\chi_1\ -g\\
\lambda^*\ \ \ \omega_2\ \ \ -g\  -2\chi_2\\
2\chi_1^*\ \ g^*\ \ -\omega_1\ -\lambda^*\\
g^*\ \ \ 2\chi_2^*\ -\lambda\ \ -\omega_2
  \end{array}
\right )  .\tag{17}\label{eq:17}\]

Firstly, we consider that $\chi_1=\chi_2=g=0$. In other words, there are only rotating wave interactions. The eigenvalues of the HB matrix are given by $\Omega_1=\frac{\omega_1+\omega_2}{2}+\sqrt{(\frac{\omega_1-\omega_2}{2})^2+|\lambda|^2}$ and $\Omega_2=\frac{\omega_1+\omega_2}{2}-\sqrt{(\frac{\omega_1-\omega_2}{2})^2+|\lambda|^2}$. When $|\lambda|=\sqrt{\omega_1\omega_2}$, $\Omega_2=0$ denotes that the DP occurs. And the eigenvalues $\Omega_1$ and $\Omega_2$ are still real, meaning that the EP never appears. The ground state is $|\psi_2\rangle=-\lambda|10\rangle+[\frac{\omega_1-\omega_2}{2}+\sqrt{|\lambda|^2+(\frac{\omega_1-\omega_2}{2})^2}]|01\rangle$. The order parameters $\langle\psi_2| a_1|\psi_2\rangle$ and $\langle\psi_2| a_2|\psi_2\rangle$ are always equal to 0 whether $|\lambda|$ is greater than or less than $\sqrt{\omega_1\omega_2}$. It shows that no quantum phase transition occurs. That means the DP is not a critical point.

Then we consider that two cavity modes couple with each other by the resonant interaction as shown in Fig.~\ref{fig.1}. The two cavity modes are generated through two crystal with second-order nonlinearity (OPA)\cite{lab27,lab28,lab29}. The Hamiltonian of two cavity systems is described as
\begin{align}
H_c&=\omega_{c1} a^\dagger_1 a_1+\omega_{c2} a^\dagger_2 a_2+[\chi_1 \exp(2i\omega_{d}t)a^2_1\nonumber\\
&+\chi_2\exp(2i\omega_{d}t) a^2_2+\lambda a_1a_2^\dagger+h.c.],\tag{18}\label{eq:18}
\end{align}
In the rotating frame, by defining the detunings of the two cavity modes as $\omega_1=\omega_{c1}-\omega_{d}$ and $\omega_2=\omega_{c2}-\omega_{d}$, the Eq.~(\ref{eq:16}) with $g=0$ is obtained. The eigenvalues of the HB matrix can be analytically achieved.

In the case of the perfect symmetry ( $\omega_1=\omega_2,\ \chi_1=\chi_2,\ \lambda=\pm|\lambda|$), the eigenvalues $\Omega_1=\sqrt{(\omega_1+\lambda)^2-4\chi_1^2}$ and $\Omega_2=\sqrt{(\omega_1-\lambda)^2-4\chi_1^2}$. The critical point of phase transition appears at $4\chi_1^2=\min\{(\omega_1+\lambda)^2,\ (\omega_1-\lambda)^2\}$. This shows that the DP and the EP coincide in the two-mode bosonic system. We verify it in multi-mode bosonic systems with the perfect symmetry ( swapping any two bosonic subsystem and keeping the total Hamiltonian invariable). It means that the value of the determinant of the HB matrix equal to 0 can be used to determine the occurrence of the phase transition in the perfectly symmetric system.

In the case of $\omega_1=\omega_2=\omega>0,\ \chi_1>0,\ \chi_2=0,\ \lambda=|\lambda|$, the eigenvalues can be calculated as
\begin{align}
\Omega_1=\sqrt{\omega^2+\lambda^2-2\chi_1^2-2\sqrt{\chi_1^4-\chi_1^2\lambda^2+\omega^2\lambda^2}},\tag{19}\label{eq:19}\\
\Omega_2=\sqrt{\omega^2+\lambda^2-2\chi_1^2+2\sqrt{\chi_1^4-\chi_1^2\lambda^2+\omega^2\lambda^2}},\tag{20}\label{eq:20}
\end{align}

For $\lambda^2>5.40205\omega^2$, the critical two-photon driving strength is given by
\begin{align}
\chi_1=\{\sqrt{\frac{\lambda^2}{2}\pm\frac{1}{2}\sqrt{1-\frac{4\omega^2}{\lambda^2}}},\ \frac{\lambda^2-\omega^2}{2\omega}\}.\tag{21}\label{eq:21}
\end{align}
When the two-photon driving strength $\chi_1=\sqrt{\frac{\lambda^2}{2}\pm\frac{1}{2}\sqrt{1-\frac{4\omega^2}{\lambda^2}}}$, the EPs show up and the DP doesn't. It shows that the DP cannot be used as a basis for judging the existence of phase transition. When $\chi_1<\sqrt{\frac{\lambda^2}{2}-\frac{1}{2}\sqrt{1-\frac{4\omega^2}{\lambda^2}}}$ or $ \frac{\lambda^2-\omega^2}{2\omega}>\chi_1>\sqrt{\frac{\lambda^2}{2}+\frac{1}{2}\sqrt{1-\frac{4\omega^2}{\lambda^2}}}$, the system is in the NP. As a conventional result\cite{lab34a}, the system will transform from the NP to the SP when the two-photon driving strength $\chi_1$ increases to $\sqrt{\frac{\lambda^2}{2}-\frac{1}{2}\sqrt{1-\frac{4\omega^2}{\lambda^2}}}$. A very interesting result is that the system will transform from the SP to the NP when the two-photon driving strength $\chi_1$ increases to $\sqrt{\frac{\lambda^2}{2}+\frac{1}{2}\sqrt{1-\frac{4\omega^2}{\lambda^2}}}$. When the two-photon driving strength $\chi_1$ increases to $\frac{\lambda^2-\omega^2}{2\omega}$ (EP conciding with DP), the system will transform from the NP to the SP again. The discovery of additional EPs will assist in the design of sensitive measuring instruments, which makes sense in quantum metrology.

\begin{figure}
\includegraphics[scale=0.85]{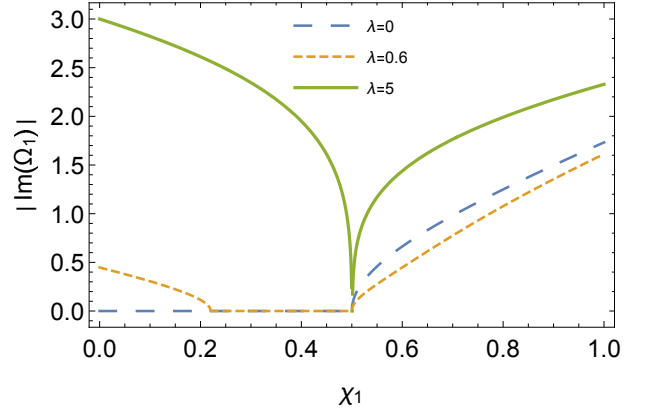}
\caption{\label{fig.2}The diagram of the absolute value of the imaginary part of the eigenvalue $\Omega_1$ changing with the two-photon driving strength $\chi_1$. Here, the value of $\omega$ is chosen to be 1.}
\end{figure}

Next, we consider the case that the counterrotating-wave interaction cannot be negligible ($g\neq0$). In the case of $\omega_1=\omega_2=\omega>0,\ \chi_1>0,\ \chi_2=0,\ \lambda=g=|\lambda|$, the Hamiltonian can be described as
\begin{align}
H_2'=\omega a^\dagger_1 a_1+\omega a^\dagger_2 a_2+(\chi_1 a^2_1+\lambda a_1a_2+\lambda a_1a_2^\dagger+h.c.).\tag{22}\label{eq:22}
\end{align}
By a similar calculation, the eigenvalues of above Hamiltonian can be achieved
\begin{align}
\Omega_1=\sqrt{\omega^2-2\chi_1^2-2\sqrt{\chi_1^4-2\chi_1\omega\lambda^2+\omega^2\lambda^2}},\tag{23}\label{eq:23}\\
\Omega_2=\sqrt{\omega^2-2\chi_1^2+2\sqrt{\chi_1^4-2\chi_1\omega\lambda^2+\omega^2\lambda^2}},\tag{24}\label{eq:24}
\end{align}
When the eigenvalue $\Omega_1$ is imaginary, the system is in the SP.  As shown in Fig.~\ref{fig.2}, we calculate the imaginary part of the eigenvalue $\Omega_1$. When $|Im(\Omega_1)|$ is nonzero, the system is in the SP; When $|Im(\Omega_1)|$ is zero, the system is in the NP. In the case of $\lambda=0$, the system transforms from the NP to the SP with the increase of the two-photon driving strength. In the case of $\lambda=5$ or $\lambda=0.6$, the systems are in the SP for the two-photon driving strength $\chi_1=0$, which is different from the previous results (without the counterrotating wave interaction). It shows that both the counterrotating-wave interaction and the two-photon driving transform the system into the SP. For a proper value of coupling strength $\lambda$ (such as, $\lambda=0.6$), the system transforms from the SP to the NP and then from the NP to the SP with the increase of $\chi_1$. Like the previous case, it is due to that the EP and the DP  don't always coincide.

\begin{figure}
\includegraphics[scale=0.55]{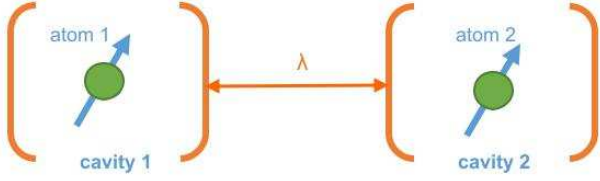}
\caption{\label{fig.3}The scheme of two-mode quantum Rabi system in cavity QED setup. The quantum Rabi subsystem composed of a two-level atom and a cavity. The two cavity modes interact with each other by the rotating-wave interaction: $\lambda a_1a_2^\dagger+h.c$.  }
\end{figure}

\begin{figure}
\includegraphics[scale=0.35]{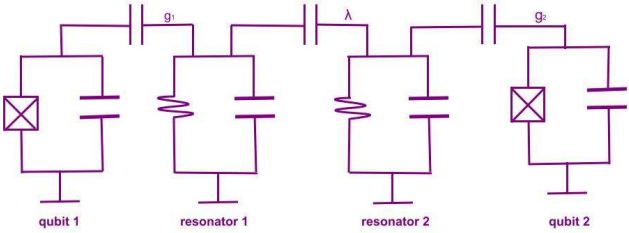}
\caption{\label{fig.4}The scheme of two-mode quantum Rabi systems in circuit QED setup.  The quantum Rabi subsystem composed of a transmon qubit and  a LC circuit or a superconductive transmission line.
}
\end{figure}

Now let's consider a system composed of two quantum Rabi subsystems, which can be realized in cavity or circuit QED\cite{lab30,lab31,lab32} as depicted in Fig.~\ref{fig.3} and Fig.~\ref{fig.4}. The corresponding Hamiltonian is described as
\begin{align}
H_{2R}=\sum_{n=1}^2[\omega_n' a^\dagger_n a_n+\frac{\Delta_n}{2}\sigma_{zn}+g_n\sigma_{xn}( a_n+a_n^\dagger)]\nonumber\\
+(\lambda a_1a_2^\dagger+h.c.).\tag{25}\label{eq:25}
\end{align}
 In the limit $\omega_n/\Delta_n\rightarrow0$ for $n=1,2$, using a Schrieffer-Wolff transformation\cite{lab26} and projecting onto the ground state,
\begin{align}
H_{2R}'=\sum_{n=1}^2[(\omega_n'-\frac{g_n^2}{2\Delta_n})a^\dagger_n a_n-\frac{g_n^2}{4\Delta_n}(a^2_n+{a^{\dagger2}_n})]\nonumber\\
+(\lambda a_1a_2^\dagger+h.c.)\tag{26}\label{eq:26}.
\end{align}
By defining $\omega_n=\omega_n'-\frac{g_n^2}{2\Delta_n}$ and $\chi_n=-\frac{g_n^2}{4\Delta_n}$ with $n=\{1,2\}$, the Eq.~(\ref{eq:16}) with $g=0$ is also obtained.
Different from the case of OPA,  $\chi_n$ and $\omega_n$ are not independent. As a result, the system transforms from the NP to the SP with the increase of $\chi_1$ in the case of $\chi_2=0$. It's not like the case of OPA where we don't get the phase transition from the SP to the NP. And when $|\lambda|^2>\omega_1'\omega_2'$, the critical point appears at the EP instead of the DP.

\section{three-mode quantum Rabi system }
In this section, we obtain the HB matrix corresponding to the three-mode quantum Rabi system and show that the quantum Fisher information will be divergent at the EP.

We consider the three-mode quantum Rabi sysytem as shown in Ref.~\cite{lab17}, which is described as
\begin{align}
H_{3R}&=\sum_{n=1}^3[\omega a^\dagger_n a_n+\frac{\Delta}{2}\sigma_{zn}+g\sigma_{xn}( a_n+a_n^\dagger)]\nonumber\\
&+\sum_{n=1,n'=1}^3J(e^{i\theta} a_na_{n'}^\dagger+e^{-i\theta}a_{n'}a_n^\dagger),\tag{27}\label{eq:27}
\end{align}
where we consider $\omega>2J>0$ for later discussion.
Implementing the Schrieffer-Wolff transformation in the limit of $\omega/\Delta\longrightarrow0$ and taking a discrete Fourier transform $a_n=\frac{1}{3}\sum_qe^{-inq}a_q$ with the quasi-momentum $q=\{0,\pm\frac{2\pi}{3}\}$, the reduced Hamiltonian is given by
\begin{align}
H_{3R}'&=\sum_{q}[\omega_q a^\dagger_q a_q-\frac{g^2}{\Delta}(a_qa_{-q}+a_q^\dagger a_{-q}^\dagger)]+E_0,\tag{28}\label{eq:28}
\end{align}
where $E_0$ is a constant and $\omega_q=\omega-\frac{2g^2}{\Delta}+2J\cos(\theta-q)$.
The eigenvalues of the HB matrix are given by
\begin{align}
\Omega_q=\frac{1}{2}[\sqrt{(\omega_q+\omega_{-q})^2-16\frac{g^4}{\Delta^2}}+\omega_q-\omega_{-q}].\tag{29}\label{eq:29}
\end{align}

In the case of $\pi/2<\theta\leq\pi$, the EP and the DP appear at $g=(\sqrt{\omega+2J\cos\theta})\Delta/2$. In the case of $0\leq\theta<\pi/2$, the EP appears at $g=(\sqrt{\omega-J\cos\theta})\Delta/2$, which is not the DP. The ground state is given by
\begin{align}
|\psi_3\rangle=\exp[\xi_0(a_0^{\dagger2}-a_0^2)+\xi_\vartheta(a_\vartheta^\dagger a_{-\vartheta}^\dagger-a_\vartheta a_{-\vartheta})]|0\rangle,\nonumber
\end{align}
where $\xi_0=\frac{1}{8}\ln\frac{\omega+2J\cos\theta}{\omega+2J\cos\theta-4g^2/\Delta}$ and $\xi_\vartheta=\frac{1}{4}\ln\frac{\omega-J\cos\theta}{\omega-J\cos\theta-4g^2/\Delta}$ with $\vartheta=\frac{2\pi}{3}$.
When close to the EP (for $\pi/2<\theta\leq\pi$, $\omega+2J\cos\theta-4g^2/\Delta=\varepsilon\rightarrow0$; for $0<\theta\leq\pi/2$, $\omega-J\cos\theta-4g^2/\Delta=\varepsilon\rightarrow0$), the dominant term of the quantum Fisher information about the parameter $\omega$ is
\begin{align}
\mathcal{F_\omega}\sim\frac{1}{32\varepsilon^2},\ \textmd{for}\ \pi/2<\theta\leq\pi;\tag{30}\label{eq:30}\\
\mathcal{F_\omega}\sim\frac{1}{16\varepsilon^2},\ \textmd{for}\ 0\leq\theta<\pi/2; \tag{31}\label{eq:31}\\
\mathcal{F_\omega}\sim\frac{3}{32\varepsilon^2},\ \textmd{for}\ \theta=\pi/2.\tag{32}\label{eq:32}
\end{align}
From above equations, we can see that the quantum Fisher information will be divergent at the EP, which reveals the quantum phase transition. $\theta=\pi/2$ can obtain the optimal estimation precision. Our results show that the EPs in the HB matrix can help to find an effective way to improve the parameter estimation precision.

\section{conclusion and outlook}
We achieve the general HB matrix for linear coupled bosonic systems in arbitrary dimensions. The EP of the HB matrix can reveal the quantum phase transition between the NP and the SP. In the single-mode bosonic or perfectly symmetric system, the DP can be the critical point due to that it coincides with the EP.
In more general multi-mode systems, the EPs and the DPs are often not coincident. As a result, unconventional and meaningful results are obtained. With the increase of the two-photon driving strength, the phase undergoes the process of \textit{NP $\rightarrow$ SP $\rightarrow$ NP $\rightarrow$ SP} in the case of neglecting the counterrotating-wave interaction. With the counterrotating-wave interaction, the process of \textit{SP $\rightarrow$ NP $\rightarrow$ SP} can be achieved. In addition, we apply the HB martrix into the quantum Rabi system and show that the quantum Fisher information will be divergent at EPs, which will lay the foundation for the design of precision measurement.
It will be interesting to further explore the dissipation transition in open system with the HB matrix.

The quantum Rabi model in this article can be realized in a variety of quantum systems, such as cold atoms\cite{lab33}, and superconducting qubits\cite{lab34}.
The form of the total Hamiltonian can be obtained by a periodic modulation of the photon hopping strength between cavities\cite{lab17}. And the strengths of two-photon
driving can be changed by the pump field and the size of the crystal, which is feasible in experiment\cite{lab24}.
\section*{Acknowledgements}
We acknowledge Fengxiao Sun for helpful discussion and constructive comments on the manuscript.
This research was supported by the National Natural Science Foundation of China under Grant No. 62001134, Guangxi Natural Science Foundation under Grant No. 2020GXNSFAA159047 and National Key R\&D Program of China under Grant No. 2018YFB1601402-2.

\end{document}